\documentclass{PoS}
\newcommand{\heading}[1]{\vspace{0.5\baselineskip}\noindent\textbf{#1}}
\newcommand{\etal}{\mbox{et al.}}
\usepackage{url}
\usepackage{tikz}
\def\checkmark{\tikz\fill[scale=0.4](0,.35) -- (.25,0) -- (1,.7) -- (.25,.15) -- cycle;} 

\title{Astrophysical Multimessenger Observatory Network (AMON): Science, Infrastructure, and Status}

\ShortTitle{AMON: Science, Infrastructure, and Status}

\author{\speaker{Azadeh Keivani}, Hugo Ayala, and James DeLaunay for the AMON core team\thanks{The AMON team URL: http://sites.psu.edu/amon/amon-participants/}\\
        The Pennsylvania State University, University Park, PA 16802, USA\\
        E-mails: \email{keivani@psu.edu}, \email{hza53@psu.edu}, \email{jjd330@psu.edu}}

\abstract{The realization of multimessenger astrophysics will open new vistas upon the most energetic events in the universe. Messenger particles of all four of nature's fundamental forces, recorded by detectors on the ground and satellites in space, enable coincidence searches for multimessenger phenomena that will allow us to discover, observe, and explore these sources. The Astrophysical Multimessenger Observatory Network (AMON) links multiple high-energy neutrino, cosmic ray, and gamma-ray observatories as well as gravitational wave facilities into a single virtual system, enabling near real-time coincidence searches for multimessenger astrophysical transients and their electromagnetic counterparts, and providing alerts to follow-up observatories. The science case, design elements, partner observatories, and status of the AMON project are presented, followed by recent results from AMON real-time and archival analyses.}

\FullConference{35th International Cosmic Ray Conference,\\
		12-20 July, 2017\\
		Bexco, Busan, Korea}

\begin{document}
\section{Introduction}

Multimessenger particles of all four of nature's fundamental forces each have the capability to provide distinct and valuable information on the most violent phenomena in the universe. Neutrinos, cosmic rays, gamma-rays (and other electromagnetic signals), and gravitational waves are presently being detected by observatories on the ground and satellites in space. Simultaneous study of these particles may help us answer fundamental questions in high-energy astrophysics, including the sources of high-energy cosmic rays and cosmic neutrinos. 

The Astrophysical Multimessenger Observatory Network (AMON) \cite{amon13} will integrate heterogeneous data from several of the world's leading high-energy observatories into a single virtual system with combined sensitivity in certain regimes greater than that of any individual facility. AMON is designed to perform coincidence searches between the sub-threshold data of different streams in real-time and distribute prompt alerts to follow-up observatories in order to exploit the multimessenger sources which are expected to be transients. 
In addition, AMON stores events from participating observatories into its database to perform archival searches for significant coincidences. AMON also receives events and broadcasts them immediately to the astronomical community for follow-up. 

In this paper, we present the AMON infrastructure as well as its participating observatories in \S\ref{sec:infra}, the current status including archival analyses and real-time searches in \S\ref{sec:stat}, and finally conclude in \S\ref{sec:conc}.


\section{Network and Infrastructure}
\label{sec:infra}
\subsection{Network}

AMON is constituted as a network of observatories and a cyberinfrastructure system for joint analysis of multimessenger data and electronic alert distribution. Data shared across the network remain the property of the originating collaborations and all decisions about data analyses and publications are made by the participating collaborations. Further information about AMON data sharing policies can be found in AMON Memorandum of Understanding (MoU)~\cite{mou}. 

\begin{figure}
\centering
\includegraphics[width=.5\textwidth]{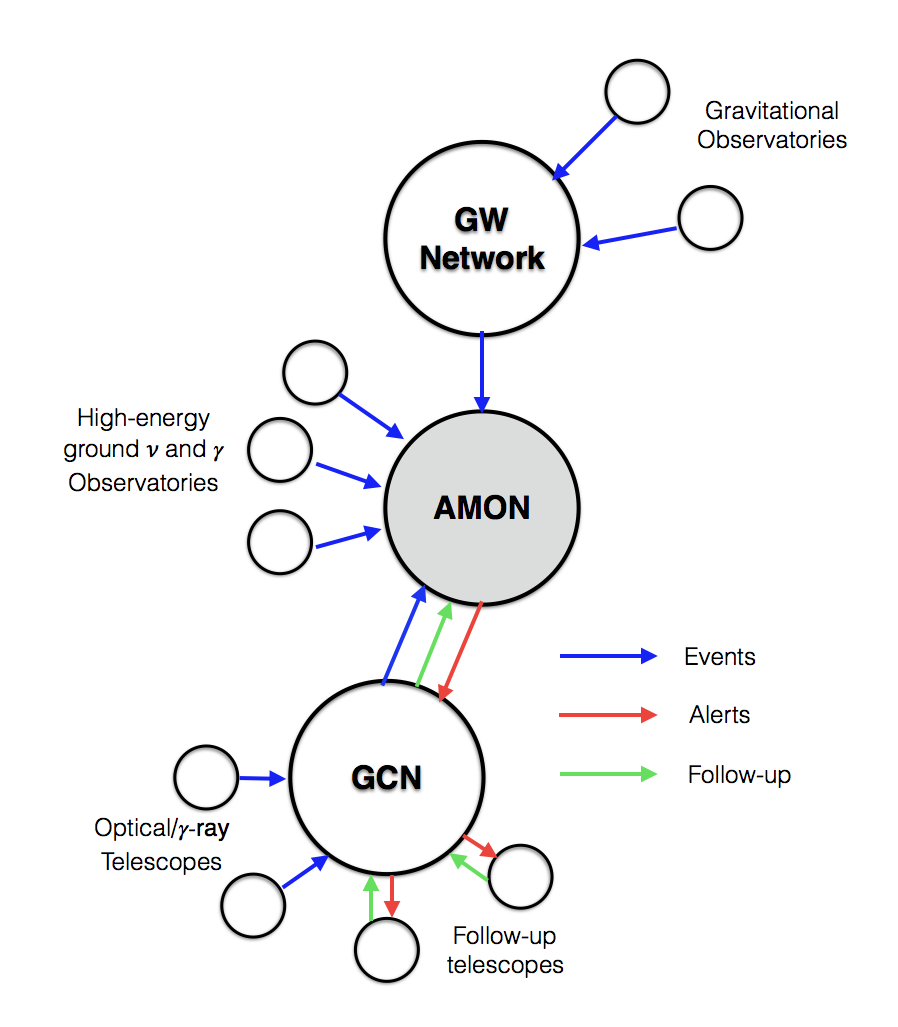}
\caption{Sub-threshold data from multimessenger triggering observatories and satellite experiments are sent to AMON.
Data from satellite experiments are sent via GCN and data from gravitational observatories are sent through the gravitational wave (GW) network. 
AMON uses GCN to distribute alerts to the follow-up observatories. Interesting follow-ups are sent back to AMON.}
\label{fig_flow}
\end{figure}

The AMON network consists of triggering and follow-up observatories. 
The triggering observatories are sensitive to one or more messenger particles and typically have large fields of view (FoV), as well as high duty cycles. Ground based triggering observatories send their candidate events to AMON via direct links in real-time. Events from satellite experiments (e.g., Swift BAT and Fermi LAT) are received by utilizing the Gamma-ray Coordinates Network (GCN)~\cite{gcn95} or directly from their websites if data are not available through GCN.
AMON also utilizes GCN to distribute its electronic alerts (in the VOEvent format~\cite{voevent}) to the follow-up observatories in real-time. The follow-up observatories are pointing telescopes situated on Earth (e.g., VERITAS, MASTER, etc.), as well as orbital telescopes (Swift XRT \& UVOT) that respond in near real-time to AMON alerts and perform follow-up observations for electromagnetic counterparts of the multimessenger signal. If the follow-up campaign results in an interesting observation, the event is sent back to AMON. 
The flow of data within the AMON network is illustrated in Figure \ref{fig_flow}. 

The partner observatories that signed the AMON Memorandum of Understanding (MoU) include the IceCube and ANTARES Neutrino Observatories, the FACT, HAWC, and VERITAS $\gamma$-ray observatories, the Pierre Auger Cosmic Ray Observatory, the Swift and Fermi satellite experiments and the MASTER robotic telescope network. Table~\ref{tab1} shows the messenger type and role for each partner observatory. 

AMON is an open and extensible network that continues to welcome new observatories and collaborators interested in the AMON scientific goals. Prospective collaborators are encouraged to contact the authors for more information about how to join the AMON network~\cite{join-amon}. 

\begin{table}
\begin{center}
\begin{tabular}{llllll}
\hline
\hline
Observatory & Messenger type & Role\\
\hline
ANTARES & neutrinos & triggering\\
FACT & gamma-rays & triggering/follow-up\\
Fermi LAT & gamma-rays & triggering\\
Fermi GBM & gamma-rays & triggering\\
HAWC & gamma-rays & triggering\\
IceCube & neutrinos & triggering\\
LIGO-Virgo* & gravitational waves & triggering\\
MASTER & optical photons & follow-up\\
Pierre Auger & neutrons/neutrinos/gamma-rays & triggering\\
Swift BAT & X rays & triggering\\ 
Swift XRT & X rays & follow-up\\ 
Swift UVOT & UV/optical & follow-up\\ 
VERITAS & gamma-rays & triggering/follow-up\\
\hline
\hline
* Ongoing MoU negotiations & & &\\
\end{tabular}
\end{center}
\caption{AMON triggering and follow-up partner observatories.}
\label{tab1}
\end{table}


\subsection{Hardware}

AMON consists of two high-uptime (less than 1 hour of downtime per year) servers, located at the Pennsylvania State University.  The machines are located in physically separate facilities on the campus. 
The system is hosted and administered by the university's Institute for CyberScience. The machines, DELL Model PowerEdge R630 servers with memory mirroring, are physically and cyber secure with  built-in hardware and power redundancies for high reliability. 


\subsection{Server}

The AMON real-time server is an asynchronous server written in Twisted~\cite{ref:twisted}, an event-driven network framework developed in Python. The server accepts XML messages with the VOEvent schema format, using the HTTPS post method. The server receives data from member observatories and will send alerts to follow-up facilities when coincident signals are found.

\begin{figure}
\centering
\includegraphics[width=.5\textwidth]{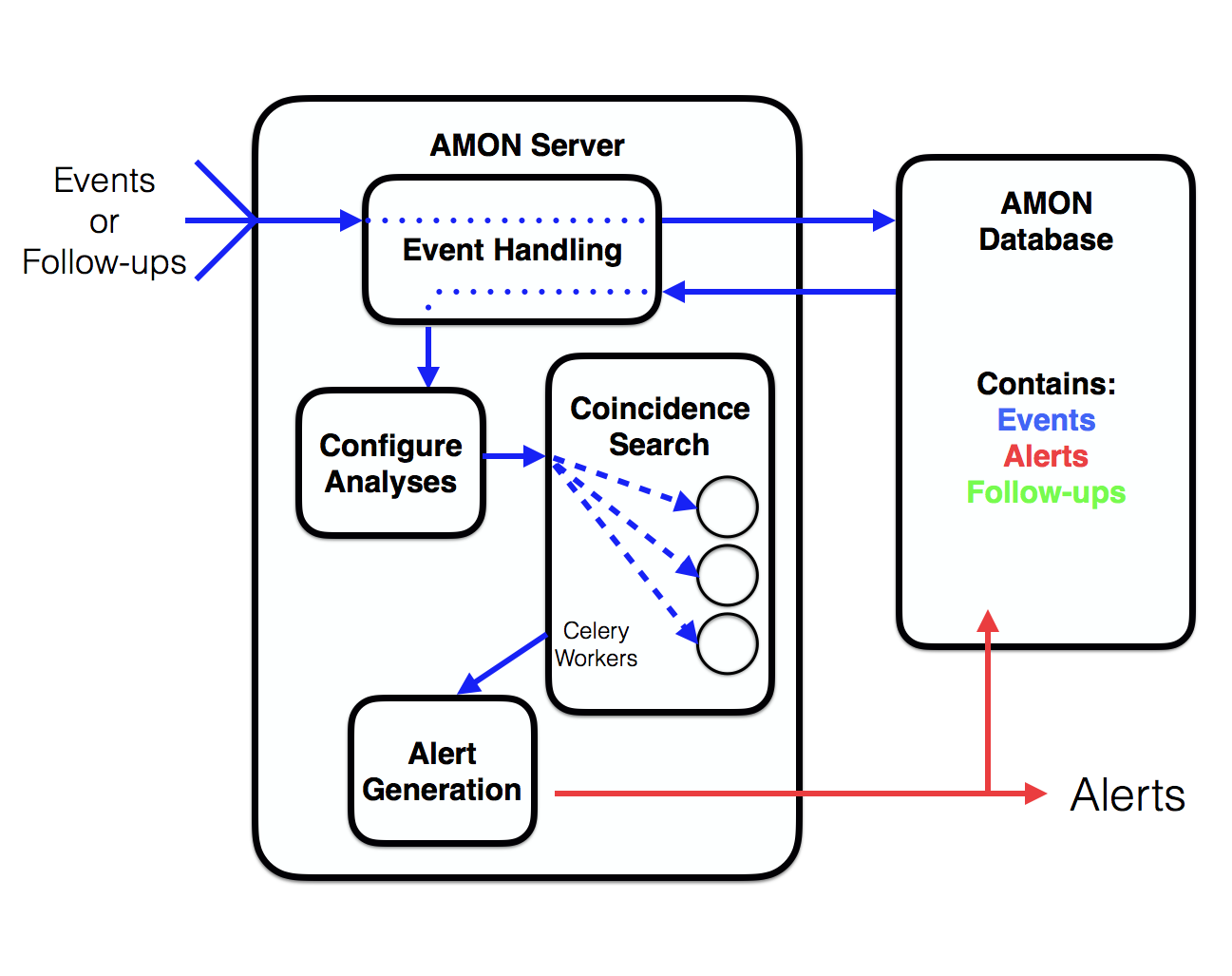}
\caption{Diagram showing the processes running in the AMON servers. Events arrive at the server asynchronously. The server writes the event into the local database and initiates the coincidence algorithm. Alerts are saved in the database and are also sent to GCN for broader (private or public) distribution.}
\label{fig:serverdiag}
\end{figure}

For each event received, the server writes it immediately to the AMON database and then sends it to be processed by the dedicated AMON real-time analysis algorithm. In order to process several events with an asynchronous order, the python software Celery~\cite{ref:celery} is used. Celery is a task queue that sends messages to different workers to do the analysis. The analysis algorithm compares each event time and arrival direction across multiple observatory streams searching for coincident events. Interesting coincidences generate AMON alerts for distribution to subscribers via GCN. Figure \ref{fig:serverdiag} shows a diagram describing the process of the AMON Server.

AMON server has been online since August 2014, processing archival data, and sending public real-time alerts since April 2016 (\S\ref{sec:realtime}). 
The event transmission software (known as the AMON client) is also written in Python/Twisted and uses the \texttt{HTTPS Post} method to send observatory events in VOEvent format to the AMON server. 
This client, as well as the entire AMON analysis software package, are made available to AMON collaborators. 


\subsection{AMON database}

The AMON system implements the world's first terabyte scale multimessenger database. The database is designed to store events received in real-time from the AMON member observatories as well as their archival data. In addition, the database stores: AMON alerts, observatory configuration models, cuts used in alert search algorithms, multiple analysis stream configurations including astrophysical sources, and information from follow-up observations of the AMON alerts. AMON's event content and alert format is described in~\cite{amon-icrc15}.

Currently, the database stores public data from: the 40- and 59-string configurations of IceCube (IC40 and IC59), as well as the first year of IceCube's 86-string configuration (IC86), Fermi LAT, and Swift BAT. 
We have also injected private archival data from Pierre Auger, one year of ANTARES (2008), and one month of HAWC data.  We should note that the database also ingests new real-time data as they are received. 
In addition, we obtained public data from LIGO runs S5 and S6 and are in the process of writing them to the database. Finally, there is an ongoing effort to get permissions from the AMON member observatories to obtain their more recent private data. 


\section{Current Status}
\label{sec:stat}

AMON's main goal is to look for coincident signals in real-time from different incoming streams. In addition to the real-time searches, we also search for significant coincident signals in the archived data. The archival studies underpin the development of future real-time coincidence analyses. 


\subsection{Archival analysis}
\label{sec:archive}

AMON archival searches are divided into a few categories, based on the type of the chosen data streams. 
These analyses may discover first multimessenger sources or provide new constrains on jointly-emitting astrophysical source populations.  In most of the following coincidence searches, alerts are pairs of events from two or more different observatories which satisfy both temporal and spatial coincidences as well as acquire high likelihood values. 


\heading{Neutrino/Gamma-Ray --- }
Many models predict that neutrino emission is paired with an electromagnetic signal (for a recent review, see \cite{kohta-review}).
Discovery of astrophysical neutrinos by the IceCube neutrino observatory~\cite{IC3yrs} has opened a new window to searches for jointly-emitting neutrino/gamma-ray transient sources. 

Within the AMON framework, we have performed coincidence analyses of the public archival data from the IceCube and Fermi LAT observatories. In one study on public neutrino data from IC40 and contemporaneous public gamma-ray data from Fermi LAT, several statistical tests on observed data using the background and signal datasets were conducted~\cite{ic-fermi-icrc15}.  However, no significant signal excess was observed.  The results of searches using IC59 and Fermi LAT, and the combined IC40/59 datasets and Fermi LAT, will be presented in future publications. The statistical tools to study correlations between IceCube neutrinos and Fermi LAT gamma-ray events have been developed and are ready to be used for further studies that include more data. 

In another study, we searched for blazar flux-correlated high-energy neutrinos from six blazars observed by VERITAS using the public IC40 data~\cite{colin16}. The lightcurves used in this search were publicly available. No significant excess of neutrinos from the preselected blazars was found in this search. Upper limits with 90\% confidence level were placed on the number of expected neutrinos from each search in this study. 

In a joint effort, AMON, IceCube, and HAWC are currently developing a correlation analysis between IceCube sub-threshold neutrinos and HAWC time-integrated daily maps, using sample of archival contemporaneous data from both IceCube and HAWC. The ultimate goal of this search is to be used in real-time searches.

Another analysis in progress searches for coincident signals between IceCube neutrinos and Swift BAT gamma-rays. Swift BAT sub-threshold data are received in AMON via GCN~\cite{swiftsubsub} to be used in the coincident alert searches. This data stream includes low significance peaks in the image domain, which are localized to $\sim$4 arcminutes in the sky and have a latency of 1--8 hours.

The AMON team is also developing a coincidence search combining IceCube sub-threshold data, Fermi GBM sub-threshold data, and Fermi Gamma-Ray Burst (GRB) data.


\heading{Neutrino/Gamma-Ray/Cosmic Ray --- }
An example of an AMON-brokered multimessenger coincidence search is the analysis of neutrino data from IceCube, gamma-ray data from HAWC and cosmic ray data from the Pierre Auger Observatory, the aim of which is to search for a distinctive primordial black hole (PBH) evaporation signature~\cite{pbh-icrc15}. Detection of PBHs would be a scientific breakthrough confirming Hawking's hypothesis of black hole radiation and cosmological models of phase transitions, and would allow us to probe physics at the highest energy scale.


\heading{Gamma-Ray/Gravitational Waves --- }
Searches for jointly emitting gamma-ray and gravitational wave sources using sub-threshold data from Swift BAT and publicly available LIGO data from runs S5 and S6~\cite{ref:ligodata} are currently under development. The LIGO data is processed with LIGO software using the GstLAL-based inspiral pipeline~\cite{ref:Cody17, ref:lal} to find "sub-threshold" triggers up to a certain false alarm rate. This pipeline can be used in a real-time setting with a latency of $\sim$1 minute, which allows for a smooth transition to a future, real-time search. 


\subsection{Real-time operations}
\label{sec:realtime}

AMON established a connection with GCN in spring 2015~\cite{gcn95, ref:scott}. The GCN client directly connects to AMON using the ``vTCP'' (VOEvent over TCP) protocol and receives AMON alerts. The AMON alert streams at GCN can be configured as private or public streams depending on the directives of the corresponding observatories whose data contribute to that particular alert stream. AMON started issuing the first real-time public alerts via GCN in April 2016 and over 50 follow-up facilities and individuals are currently subscribed to it.

At present there are four event streams from the IceCube neutrino observatory transmitted to AMON in real-time: muon neutrino ($\nu_\mu$) singlet, neutrino multiplets, high-energy starting events (HESE), and extremely high-energy (EHE) events. The last two streams have been distributed publicly through GCN since April and July 2016, respectively. The multiplet stream is currently distributed privately to IceCube's partner optical follow-up observatories as part of the IceCube optical follow-up program. The singlet stream has not yet been cleared for analysis use pending final approval from the IceCube Collaboration. This stream will be used in multiple real-time coincident searches, such as in IceCube-HAWC and IceCube-Swift BAT searches that were discussed in \S\ref{sec:archive}. The singlet stream contains sub-threshold data at the rate of about 5.0~mHz.

Public data from Swift BAT and Fermi LAT are received and stored in the AMON database in near real-time as soon as they are available through GCN or on their website. These data are available to all AMON-brokered analyses and will be used in real-time coincident analyses along with IceCube singlet neutrinos. 

Other real-time data streams that are currently under development for inclusion in AMON are sub-threshold data from the Pierre Auger Observatory and the FACT gamma-ray observatory.  Both observatories are currently sending test alerts to AMON in the form of actual data that has been scrambled.  We are awaiting approval from the Pierre Auger Observatory to unblind the stream.  AMON and FACT are currently discussing the final trigger criteria for FACT alerts. 

Additionally, there are ongoing efforts with the HAWC Collaboration to transmit its daily map data stream in real-time to AMON. The current status of real-time streams from the participating facilities is summarized in Table~\ref{tab2}.
\begin{table}
\begin{center}
\scalebox{0.9}{
\begin{tabular}{ccccccc}
\hline
\hline
Obs. & Stream  & Stream  & TLS  & Test  & Test  & Real \\
 & name & content & certificates & (simulations) & (blinded data) &  data\\
\hline
IceCube & singlet & \checkmark & \checkmark & \checkmark & \checkmark & in progress\\
IceCube & multiplet & \checkmark & \checkmark & \checkmark & \checkmark & \checkmark\\
IceCube & HESE & \checkmark & \checkmark & \checkmark & \checkmark & \checkmark\\
IceCube & EHE & \checkmark & \checkmark & \checkmark & \checkmark & \checkmark\\
Pierre Auger & sub-threshold & \checkmark & \checkmark & \checkmark & \checkmark & in progress\\
FACT & sub-threshold & \checkmark & \checkmark & \checkmark & \checkmark & in progress\\
HAWC & daily-maps & \checkmark & in progress & in progress & in progress & in progress\\
Swift BAT & sub-threshold & \checkmark & n/a &  n/a &  n/a & \checkmark\\
Fermi LAT & sub-threshold & \checkmark & n/a &  n/a &  n/a & \checkmark\\
\hline
\hline
\end{tabular}}
\end{center}
\caption{Status of AMON active real-time streams}
\label{tab2}
\end{table}  


\section{Conclusion}
\label{sec:conc}
The AMON infrastructure enables real-time and archival searches for possible multimessenger sources by conducting joint sub-threshold analysis of multimessenger data. AMON has made significant progress toward both real-time and archival analyses. AMON has been distributing IceCube alerts via GCN since April of 2016 and it will start issuing coincident alerts by the end of this year.  AMON is ready to distribute data streams publicly or to private recipients per request of its partner observatories. The AMON partner facilities jointly probe the high-energy universe via all four forces of nature to discover the sources of high-energy neutrinos and cosmic rays.


{\noindent\bf Acknowledgments}
The authors acknowledge support from the National Science Foundation under grant PHY-1412633 and the Institute for Gravitation and the Cosmos at the Pennsylvania State University.  This research or portions of this research were conducted with Advanced CyberInfrastructure computational resources provided by The Institute for CyberScience at The Pennsylvania State University (http://ics.psu.edu).


\end{document}